\documentstyle[preprint,eqsecnum,aps]{revtex}

\newcommand{\be}{\begin{equation}}
\newcommand{\ee}{\end{equation}}
\newcommand{\bea}{\begin{eqnarray}}
\newcommand{\eea}{\end{eqnarray}}
\newcommand{\bem}{\begin{mathletters}}
\newcommand{\eem}{\end{mathletters}}
\newcommand{\nn}{\nonumber}
\newcommand{\sla}{\! \not \!}
\newcommand{\otto}{\leftrightarrow}
\newcommand{\Punkt}{\quad .}
\newcommand{\Komma}{\quad ,}

\begin{document}

\draft
\preprint{SUBATECH--98--20}

\title{Hadronization and Strangeness Production in a
       Chirally Symmetric Nonequilibrium Model}

\author{P.~Rehberg and J.~Aichelin}
\address{SUBATECH \\
         Laboratoire de Physique Subatomique et des Technologies
                  Associ\'ees \\
         UMR Universit\'e de Nantes, IN2P3/CNRS, Ecole des Mines de Nantes \\
         4 Rue Alfred Kastler, F-44070 Nantes Cedex 3, France}

\maketitle
\begin{abstract}
The expansion and hadronization of a quark meson plasma is studied using
an effective chiral interaction Lagrangian. The particles we consider
are light as well as strange quarks, which can form pions, kaons and
$\eta$ mesons via collision processes. The transport equations for the
system are solved using a QMD type algorithm. We find that in chemical
equilibrium at high temperatures the strange quark mass is considerably
higher than the strange current quark mass and becomes even higher
if we assume an initial state free of strange quarks. This leads to
a considerably higher production threshold. In contrast to simpler
scenarios, like thermodynamics of free quarks with their bare mass,
we observe that strangeness production in a plasma is hindered and not
favoured. The different particle species created during the evolution
become separated in coordinate as well as in momentum space. We observe,
as at CERN experiments, a larger mean momentum of kaons as compared to
pions. Thus the radial collective velocity may as well originate from
a plasma expansion and not necessarily from a hadronic scenario.
\end{abstract}

%
%

\clearpage

\section{Introduction} \label{introsec}
The production of strange particles in relativistic heavy ion collisions
has received lots of attention since it was proposed that their enhanced
production could serve as an experimental signal for the creation of
a quark-gluon plasma (QGP) \cite{KMR}. In fact, it has been observed
in sulfur--sulfur and lead--lead collisions at CERN that more strange
particles have been produced than expected from an extrapolation of
proton-proton or proton--nucleus collisions. Presently all cascade
or string models fail to explain this enhancement in a hadronic
scenario. Purely phenomenological models predict in addition that the
disintegration of the plasma leads to the distillation of strangeness
or to the formation of strangelets, i.\,e. droplets of multistrange
quark matter which are stable for a sufficiently long time to be
detected. Several experiments at Brookhaven and CERN are devoted to the
search of such strangelets.

Therefore it is tempting to see whether approaches more closely related
to quantum chromodynamics (QCD) can confirm strangeness enhancement
if during the heavy ion collision a quark gluon plasma is formed. One
of the major drawbacks in the interpretation of the experimental
data is the lack of a theoretical model which is able to describe the
formation, hadronization and decay of a QGP. An ideal model for a heavy
ion collision should in principle be able to describe both quark and
hadronic matter as well as the transition between these two regimes. Due
to the expected short lifetime of a QGP, it should further be able to
handle nonequilibrium effects. A theory accomplishing all of this
is, however, unavailable nowadays, since the mechanisms leading to
confinement are presently not understood. Although QCD is known via
lattice calculations to contain this effect, a phenomenologically useful
solution of this theory does up to now not exist, so that it is impossible
to apply QCD to all stages of a heavy ion collision. This inhibits the
construction of a theory which is able to explain all phenomena. It is
nevertheless possible to construct nonequilibrium scenarios for effective
interactions, which describe at least a part of the observed phenomena.

Especially for the low energy sector of strong interactions, it has
been known for a long time that confinement does not play an important
role. This sector is rather governed by chiral symmetry. One of the many
effective chiral interaction Lagrangians known is the Nambu--Jona-Lasinio
(NJL) model \cite{nambu,sandi}, which starts from a Lagrangian containing
quarks interacting via a four point coupling. This interaction leads
to a spontaneous breakdown of chiral symmetry and to the appearance of
light pseudoscalar mesons, which are bound states of quark-antiquark
pairs. In the vacuum, it has been successfully applied to the computation
of static quantities like hadronic mass spectra \cite{sandi,bochum}
or dynamic quantities like pion scattering lengths \cite{quak}, to give
only two examples. In the medium, it has been demonstrated that chiral
symmetry becomes restored at sufficiently high temperatures and/or
chemical potentials. In this case, the effective quark masses drop
down, whereas the mesons cease to be bound states and become unstable
resonances \cite{gerry}. This effect models, to a certain extent, the
deconfinement transition of QCD. In fact it can be seen by comparing
the NJL mass spectra with lattice computations, as have been shown in
Ref.~\cite{edwin}, that the NJL model provides at least a qualitatively
correct picture of strong interactions even beyond the chiral phase
transition. The drawbacks of this model, however, are that it is
nonrenormalizable and does not confine, so that one has free quarks at
all temperatures.

Beside these successful applications to the equilibrium theory of
strong interactions, the NJL model has recently been developed further
towards a nonequilibrium formalism \cite{zhawi,trap,pion}. First
numerical calculations within this formalism have been reported in
Refs.~\cite{trap,abada,bot,bitch}. The advantages of this approach
compared to other effective models are obvious: Since one has both
quark and hadronic degrees of freedom, where the latter appear, as in
QCD, as bound states, one is, at least in principle, able to model a
transition from an initial state, which contains only quarks, to a
hadronic state. Since the numerical calculations needed turn out to be
rather time consuming, the simplicity of the NJL model is another
pluspoint.

The numerical method for the solution of the transport equations
employed here is an algorithm of the quantum molecular dynamics (QMD)
type \cite{qmd}, which means that we parametrize the Wigner function
as a sum over double gaussians and solve the equations of motion
for the parameters. This method has been chosen previously in the
NJL model in Refs.~\cite{bot,bitch}. The present paper is a follow
up of Ref.~\cite{bitch}, where we studied the two flavor version of
the NJL model. Here we extend this work in including a third quark
flavor, which enables us to study the production mechanisms of strange
particles. Several results obtained in the framework of this model,
which are not specific to three flavor calculations, have been reported
in \cite{bitch} and will not be repeated here.

One of the major results we find here is that chiral symmetry predicts
a much higher effective strange quark mass than the current mass of
ca.~150\,MeV. While this is the case already in equilibrium, where we
find a strange quark mass of 300\,MeV at a temperature of 350\,MeV,
the strange quark mass gets even further enhanced if one applies initial
conditions which do not contain strange quarks. This raises the question,
if the original models for strangeness enhancement, which are based on
an assumed mass drop of the strange quark down to the current mass, have
not been too naive. Unfortunately, since this is a generic nonequilibrium
effect, it is not possible to confirm it via lattice simulations.

This paper is organized as follows:  In Sec.~\ref{modsec}, we briefly
review the three flavor NJL model and describe our numerical algorithm.
Numerical results are presented in Sec.~\ref{numsec}. Section~\ref{sumsec}
contains our summary and conclusions.

\section{Description of the Model and the Algorithm} \label{modsec}
\subsection{The Model}
The model we use throughout this paper is the Nambu--Jona-Lasinio (NJL)
model \cite{nambu} in its three flavor version. This model is defined by
the Lagrangian
\bea
{\cal L} &=& \sum_{f=u,d,s} \bar\psi\left(i\sla\partial - m_{0f}\right)\psi
+G \sum_{a=0}^8\left[\left(\bar\psi\lambda_a\psi\right)^2
                   + \left(\bar\psi i\gamma_5\lambda_a\psi\right)^2\right]
\label{lagra} \\ \nn
&+& K\left[\det \bar\psi\left(1+\gamma_5\right) \psi
         + \det \bar\psi\left(1-\gamma_5\right) \psi \right]
\Punkt
\eea
Here, $\psi$ denotes the quark fields, which are implicitly understood
to carry flavor and color indices. The matrices $\lambda_a$ are the
Gell-Mann matrices in flavor space for $a=1,\dots,8$ and $\lambda_0 =
\sqrt{2/3}\,{\bf 1}$.  A small explicit chiral symmetry breaking is
provided by the current quark masses $m_{0f}$. $G$ and $K$ are coupling
constants with dimensions MeV$^{-2}$ and MeV$^{-5}$, respectively. The
't\,Hooft determinant proportional to $K$ serves to model the $U_A(1)$
symmetry breaking, which in QCD takes place by an effective interaction
of $2N_f$ quarks due to instanton effects.

The properties of the Lagrangian (\ref{lagra}) have been extensively
studied in the literature. We thus review here only briefly those
topics of the NJL model, which are essential for the understanding of
the present article and refer for more details to Ref.~\cite{sandi}.
The most important feature of the interaction (\ref{lagra}) is that it
preserves the chiral symmetry of QCD, i.\,e.  in the limit $m_{0f}\to 0$
it is invariant under transformations of the form
\be
\psi \to \exp\left(-i\gamma_5\sum_{a=1}^8\theta_a\lambda_a\right) \psi
\ee
for arbitrary real $\theta_a$. In the vacuum, this symmetry is
spontaneously broken and the quarks obtain an effective mass. As a
consequence of the Goldstone theorem, eight massless modes appear
as bound states in the quark-antiquark scattering matrix. These
massless modes carry the quantum numbers of the light pseudoscalar
mesons, i.\,e. one obtains three pions, four kaons and one $\eta$ meson.
The $\eta'$ meson, which appears also as a pole in the quark-antiquark
scattering matrix, is massive due to the $U_A(1)$ breaking by the
't\,Hooft determinant. For finite current quark masses, $m_{0f}\ne 0$,
this picture changes slightly in that chiral symmetry is no longer
an exact symmetry. This leads to finite meson masses, which, by
parameter choice, can be adjusted to the experimentally observed
values.

At sufficiently large temperatures and/or chemical potentials, chiral
symmetry becomes restored. This means that in this region of the
phase diagram the quark mass is either zero for $m_{0f}=0$ or
at least low for $m_{0f}\ne 0$. The mesons, on the other hand, are
no longer bound states but become resonant states, which have a finite
width due to the possible decay channel $M\to q\bar q$.

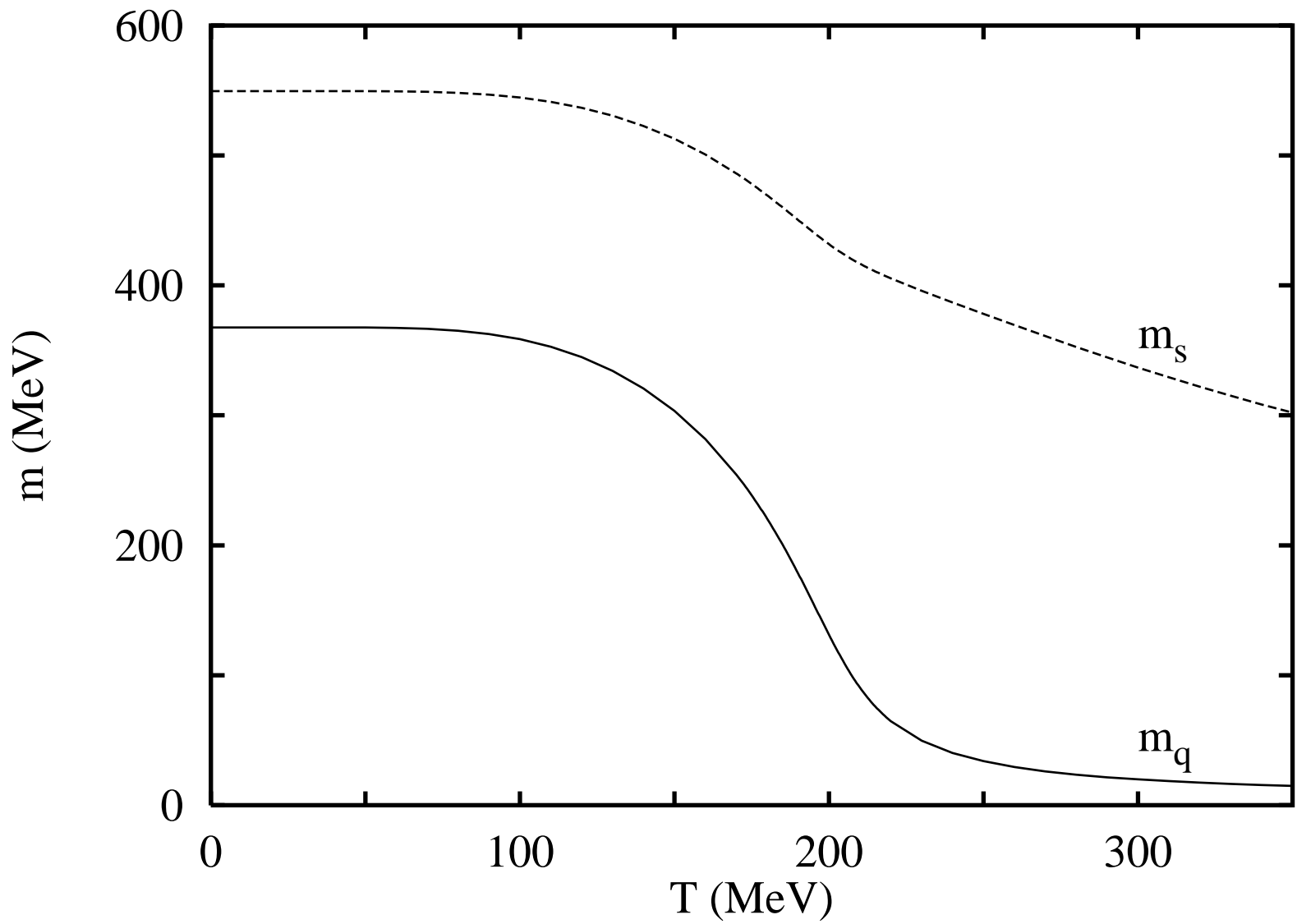
\begin{figure}
\caption[]{Constituent quark masses at finite temperature in
           equilibrium. Solid line: light quarks, dashed line: strange
           quarks.}
\label{qmass}
\end{figure}

To give a quantitative picture, we show in Fig.~\ref{qmass} the masses
of the constituent quarks at chemical potential $\mu=0$ as a function
of the temperature. In this figure, the masses are computed using
the Hartree approximation, which is the leading order of an expansion
in the inverse number of colors, $1/N_c$ \cite{expand}. This
leads to the gap equation
\bem \label{gap} \bea
m_i &=& m_{0i} - \frac{GN_c}{\pi^2} m_i A_i 
        + \frac{KN_c^2}{8\pi^4} m_j A_j m_k A_k \Komma
          \qquad i\ne j\ne k \ne i \\
\label{adef}
A_i &=& -8 \pi^2 \int_{\left|\vec p\right|<\Lambda} \frac{d^3p}{(2\pi)^3}
       \frac{1}{\sqrt{\vec p\,^2+m_i^2}}
       \left(1- \frac{n_i + n_{\bar i}}{2N_c}\right) \Komma
\eea \eem
where the indices $i$, $j$ and $k$ run over all three quark flavors.
The phase space distribution of quarks and antiquarks of flavor $i$
is denoted by $n_i$ and $n_{\bar i}$, respectively. In equilibrium,
they are given by the Fermi distribution function.  Since the NJL model
is nonrenormalizable, the integral in Eq.~(\ref{adef}) has been limited
to momenta smaller than an $O(3)$ cutoff $\Lambda$. The parameters
used in Fig.~\ref{qmass} are $m_{0q}=5.5$\,MeV, $m_{0s}=140.7$\,MeV,
$G\Lambda^2=1.835$, $K\Lambda^5=12.36$ and $\Lambda = 602.3$\,MeV,
where we use the generic index $q$ to denote both $u$ and $d$.  At $T=0$,
chiral symmetry is spontaneously broken and one obtains constituent quark
masses of $m_q=368$\,MeV and $m_s=550$\,MeV.  These masses stay more or
less constant up to a temperature of approximately 200\,MeV, where the
light quark mass drops down to a value close its bare value. The strange
quark mass, on the other hand, also drops down but stays relatively large.
At $T=350$\,MeV, which is far beyond any temperature to be expected in
a heavy ion experiment, one still has an effective strange quark mass of
$m_s=300$\,MeV, which is about two times the bare strange quark mass.

Mesons appear as poles in the quark-antiquark scattering matrix. A
computation of this quantity leads to the meson dispersion relation
\cite{sandi}
\be
1-2G\Pi^R_{PS}(p) = 0 \Komma \label{mesdis}
\ee
where $\Pi^R_{PS}(p)$ is the irreducible retarded pseudoscalar
polarization function. In lowest order of $1/N_c$ it is given by the
diagram shown in Fig.~\ref{polarfig}.  The temperature dependence of the
meson masses is shown in Fig.~\ref{mmass}, together with the temperature
dependence of $2m_q$ and $m_q+m_s$. The pion mass is denoted by the
solid line of Fig.~\ref{mmass}. At zero temperature, its mass is equal
to the experimental value of 135\,MeV. This mass stays roughly constant
until it begins to rise at a temperature around 200\,MeV. At the pion
Mott temperature $T_{M_\pi}=212$\,MeV, which is marked by the arrow in
Fig.~\ref{mmass}, its mass becomes equal to that of its constituents,
$m_\pi(T_{M_\pi})=2m_q(T_{M_\pi})$, and stays above at higher
temperatures.  In this temperature range, the pion becomes unstable due
to a Mott effect and obtains a finite width corresponding to the decay
channel $\pi\to q\bar q$. This Mott effect models to a certain extent the
deconfinement transition of QCD. For more details about the Mott transition
in the NJL model see Ref.~\cite{gerry}.

The kaon behaves in a similar way. At zero temperature, one obtains a mass
of 497\,MeV. At the kaon Mott temperature $T_{M_K}$ one has $m_K(T_{M_K})
= m_q(T_{M_K}) + m_s(T_{M_K})$ and the kaon becomes unstable at higher
temperatures. Numerically, one finds $T_{M_K} = 210$\,MeV$ \approx
T_{M_\pi}$. The same behaviour is found for the $\eta$ meson. Whereas one
has $m_\eta=515$\,MeV at zero temperature, one finds a Mott transition
at $T_{M_\eta}=180$\,MeV and an unstable resonance above. The $\eta'$ is
special, since due to its large mass and the lack of confinement in the NJL
model it is unstable at all temperatures. For this reason and because the
hadronization cross sections for the $\eta'$ production are comparatively
low \cite{su3hadron}, we will not consider the $\eta'$ further.

\subsection{The Simulation Algorithm}
Our simulation is based on the observation, that both quark and meson
degrees of freedom can be described {\em simultaneously \/} by
transport equations of the Boltzmann type \cite{pion},
\be \label{boltz}
\left(\partial_t + \vec\partial_p E \vec \partial_x
      - \vec\partial_x E \vec \partial_p \right) n(\vec x, \vec p, t)
    = I_{\rm coll}[n(\vec x, \vec p, t)]
\ee
where $n(\vec x, \vec p, t)$ is the Wigner function of the particle species
in question and $E$ the corresponding quasiparticle energy. This
quantity is in general a complicated function of $\vec x$, $\vec p$
and $t$, which has to be determined in a selfconsistent fashion from
a diagrammatic expansion of the self energy or the irreducible polarization,
respectively \cite{pion}. In the following, we will make the approximation
\be
E(\vec x, \vec p, t) = \sqrt{\vec p\,^2 + m(\vec x, t)^2} \Komma
\ee
where the quasiparticle mass $m(\vec x, t)$ is determined either from
Eq.~(\ref{gap}) for quarks or from the solution of Eq.~(\ref{mesdis})
with $\vec p = 0$ for mesons.

For the solution of Eq.~(\ref{boltz}) we employ an algorithm of the
QMD type \cite{qmd}. To this end, we parametrize the Wigner
function as a sum over double Gaussians,
\be \label{para}
n(\vec x, \vec p, t) = \sum_{i=1}^N
     \exp\left(-\frac{\left(\vec r - \vec r_i(t)\right)^2}{2w^2}\right)
     \exp\left(-\frac{w^2}{2} \left(\vec p - \vec p_i(t)\right)^2\right)
\Punkt
\ee
The normalization is chosen in such a way that the integral over
$n$ is equal to the total number of particles,
\be
\int \frac{d^3x \, d^3p}{(2\pi)^3} n(\vec x, \vec p, t) = N \Punkt
\ee
The centroids $r_i(t)$, $p_i(t)$ move along the characteristics of
Eq.~(\ref{boltz}), i.\,e.
\bem \label{charisma} \bea
\dot{\vec x}_i(t) &=& \vec\partial_p E(\vec x_i(t), \vec p_i(t), t) \\
\dot{\vec p}_i(t) &=& - \vec\partial_x E(\vec x_i(t), \vec p_i(t), t)
+ \mbox{collision contributions} \Punkt \label{ppunkt}
\eea \eem
The collision processes which enter in Eq.~(\ref{ppunkt}) belong to three
different classes: (i) quark elastic scattering processes, $qq\otto qq$,
$q\bar q\otto q\bar q$ and $\bar q\bar q\otto\bar q\bar q$ \cite{elaste},
(ii) hadronization processes, $q\bar q\otto MM$ \cite{su3hadron,su2hadron}
and (iii) meson decay processes $M\to q\bar q$ \cite{pion}. The first
of these two classes are treated in the following fashion: For each
particle pair, we perform a transformation to the rest system
of the pair and check, whether these two particles will have their closest
approach within the current time step, which is a necessary condition
for a collision to happen. If this condition is fulfilled, we compute
the cross sections $\sigma_k$ for each process, which is possible for
the incoming pair, and the total cross section $\sigma_{\rm tot}=\sum
\sigma_k$. A collision happens if the minimal distance of the particle
trajectories is smaller than $\sqrt{\sigma_{\rm tot}/\pi}$. The
actual collision process is chosen randomly with probability
$\sigma_k/\sigma_{\rm tot}$. Neglecting anisotropies of the differential
cross section, we choose the scattering angle randomly in the rest frame
of the collision. This scheme works if one has a scattering process
with two particles in the initial state. For the meson decay processes,
which are only possible in the early stage of the expansion, when mesons
exist as resonances, we proceed by computing the mean life time $\tau$
of the meson in question and decide with probability $1-\exp(-\Delta
t/\tau)$, where $\Delta t$ is the time step, whether the meson decays
during this iteration or not. If the decay takes place, we again choose
the momenta of the outgoing quarks randomly in the particle rest frame
and boost them to the original frame.

The numerical task to accomplish consists thus in a numerical solution
of Eq.~(\ref{charisma}) together with Eqs.~(\ref{gap}), (\ref{mesdis}),
where the particle distribution functions appearing in the two latter
equations have to be replaced by the parametrization (\ref{para}).
For the computation of the collision contributions, one has also
to compute cross sections, which themselves are complicated functionals
of the particle distribution functions \cite{su3hadron,elaste,su2hadron}.
Doing this exactly is a task which lies far beyond present days computer
capacities.  We thus decide to take a shortcut in defining effective
thermal quantities. This works as follows: at each time step, we solve
Eq.~(\ref{gap}) with Eq.~(\ref{para}) inserted for the quark masses.
Afterwards, we define an effective temperature with the help of the
equation
\be \label{tdef}
m_q(\vec x, t) = m_q^{\rm eq}\left(T_{\rm eff}(\vec x, t)\right)
\Komma
\ee
where $m_q^{\rm eq}(T)$ is the equilibrium temperature dependence of the
light quark mass. The meson masses are then computed from the equilibrium
expressions, which are functions of $T_{\rm eff}$, $m_q$ and $m_s$. Mass
gradients, which enter Eq.~(\ref{ppunkt}), are computed via an exact
differentiation of Eq.~(\ref{gap}) for quarks, whereas for mesons we take
\be
\vec\partial_x m_M = \left[ \frac{\partial m_M^{\rm eq}}{\partial T}
                     \left(\frac{dm_q^{\rm eq}}{dT}\right)^{-1}
                     + \frac{\partial m_M^{\rm eq}}{\partial m_q}
                     \right] \vec\partial_x m_q
                     + \frac{\partial m_M^{\rm eq}}{\partial m_s}
                     \vec\partial_x m_s
\ee
in agreement with our prescription for the computation of the meson
masses. At last we compute the scattering cross sections as functions
of $T_{\rm eff}$ as well as the quark and meson masses, again using
equilibrium expressions. Note that this procedure does not necessarily
give the same numbers as in thermal equilibrium, since we have {\em two\/}
independent parameters, $m_q$ and $m_s$, which in thermal equilibrium are
coupled to each other, whereas there is no unique relation between these
two quantities in our nonequilibrium calculation. Since we do not consider
baryons in our hadronization processes, we limit ourselves to systems
with zero baryon density, i.\,e. we set $\mu_q=\mu_s=0$ in all equilibrium
expressions.

As initial condition we use a spherically symmetric system with a given
radius $r_0$, in which the spatial centroids of the quark distribution
functions are distributed with uniform probability. For the momentum
centroids of the light quark distribution function, we choose a Fermi
distribution,
\be \label{plight}
P(\vec p) \sim
\frac{2N_c}{\exp\left(\sqrt{\vec p\,^2 + m_q^{\rm eq}(T_0)^2}/T_0\right) + 1}
\Theta\left(\Lambda - p\right)
\Punkt
\ee
In this distribution, $T_0$ is a free parameter. The number of light
quarks per flavor and particle/antiparticle degree of freedom is fixed as
the momentum integral over the Fermi distribution (\ref{plight}) times the
volume. The $\Theta$-factor serves to cut off the distribution function
at momenta larger than $\Lambda$. This is necessary, since due to the
cutoff in Eq.~(\ref{gap}) energy conservation is only valid strictly
if no particles with momenta larger than $\Lambda$ are present in the
system \cite{trap}. We nevertheless expect this effect not to be of great
importance, so that below we will also study initial conditions without
this factor. For the strange quarks, the momentum distribution is chosen
to be
\be \label{pstrange}
P(\vec p) \sim f_s
\frac{2N_c}{\exp\left(\sqrt{\vec p\,^2 + m_s^{\rm eq}(T_0)^2}/T_0\right) + 1}
\Theta\left(\Lambda - p\right)
\Punkt
\ee
This differs from Eq.~(\ref{plight}) by the factor $f_s$, which serves to
adjust the degree of chemical equilibration in the initial state. For
$f_s=0$, one has no strangeness at all initially, whereas for $f_s=1$
one starts with a totally equilibrated system.

\section{Numerical Results} \label{numsec}
In this section we present the numerical outcome of our calculations.
The NJL parameters we use throughout this section are $m_{0q}=5.5$\,MeV,
$m_{0s}=140.7$\,MeV, $G\Lambda^2=1.835$, $K\Lambda^5=12.36$ and
$\Lambda=602.3$\,MeV. The width of the wave packets was chosen to be
2\,fm. The initialization temperature in Eqs.~(\ref{plight}),
(\ref{pstrange}) was taken to be $T_0=280$\,MeV.

\subsection{Space--Time Dependence of Quark Masses}
The evolution of the space-time dependence of the constituent quark masses
can be seen from Fig.~\ref{massev}. Here we show the constituent quark
masses as a function of $r$ for different times. In this calculation,
no strange quarks were present in the initial state, but Eq.~(\ref{gap})
involves nevertheless also the computation of the strange quark mass.
At $t=0$, the light quark mass is low in the center, where the particle
density is high. It begins to rise near the surface due to the gaussian
shape of the parametrization (\ref{para}). Since our program computes
the constituent quark masses only at those points, where particles are
present, the plot stops at the maximal radius $r_0=7$\,fm. At later
times, the particles flow out and thus the particle density drops.
This leads to an increase of the light quark mass. At $t=25$\,fm$/c$,
one has everywhere a light quark mass equal to the vacuum mass. The
strange quark mass behaves similar, but there is one important difference
between strange and light quarks: Whereas the value of the light quark
mass in the centre is low, as one would expect from a hot system, the
strange quark mass is much higher as in the thermal case. Numerically,
one has $m_q=41$\,MeV and $m_s=464$\,MeV in the centre, whereas for
a thermal system one would have $m_q=23$\,MeV and $m_s=353$\,MeV at
$T=280$\,MeV. The discrepancy is not very large for the light quarks,
but amounts to 110\,MeV for the strange quarks. As will be seen below,
this effect has a large impact on the production of strange particles,
because the production threshold rises considerably. The reason for this
effect can be seen from Eq.~(\ref{gap}): writing this equation explicitly
for light and strange quarks, one has
\bem \label{joerggap} \bea \label{leger}
m_q &=& m_{0q} - \left(G - \frac{KN_c}{8\pi^2} m_s A_s\right)
                 \frac{N_c}{\pi^2} m_q A_q
\\ \label{etrange}
m_s &=& m_{0s} - \frac{GN_c}{\pi^2} m_s A_s 
        + \frac{KN_c^2}{8\pi^4} \left(m_q A_q\right)^2 \Punkt
\eea \eem
Note that all selfenergy contributions on the right hand side of
Eqs.~(\ref{joerggap}) increase the mass and have their maximum at
vanishing density. Since $n_s(\vec x, \vec p, t = 0) = 0$, the second term
on the right hand side of Eq.~(\ref{etrange}) does {\em not\/} receive
direct medium corrections and thus gives a {\em larger\/} contribution
to the mass as it would do in chemical equilibrium.  There are medium
corrections to the strange quark mass entering through the third term
on the right hand side of Eq.~(\ref{etrange}) which cause the mass drop
in the centre, which is seen in Fig.~\ref{massev}. However, these
contributions are not sufficient to generate a strange quark mass near
the thermal strange quark mass. In order to have the strange quarks
equilibrated, they first have to be created with a much larger mass
than the equilibrium mass, which makes it difficult for the system to
approach equilibrium.

We would like to stress that this effect can presently only be seen in
phenomenological models like the NJL model. Since it is a nonequilibrium
effect, there are no means to reproduce or to disprove it in lattice
calculations.

\subsection{Particle Multiplicities and Meson Production Mechanisms}
A first account of the meson production mechanisms in the NJL model has
been given in Ref.~\cite{bitch}, where the expansion and hadronization
of a two flavor plasma has been studied. The general results reported
there remain true for a three flavor plasma. Nevertheless, there are
some phenomena of a three flavor plasma, which do not exist in a two
flavor plasma.

One of these aspects which can be studied within our hadronization model
for a quark gluon plasma are the particle multiplicities, especially the
ratio of strange to nonstrange particles. These quantities are interesting
because it was claimed already several years ago that they could
serve as an experimental signal for the creation of a quark-gluon plasma
\cite{KMR}. Due to the shortcomings of our model, we cannot expect to
obtain precise numbers for the particle multiplicities, but we expect
to obtain a qualitatively reasonable estimate. To this end, we show in
Tab.~\ref{multab1} the multiplicities obtained from the initial conditions
(\ref{plight}), (\ref{pstrange}) for different initial radii. It can be
seen that for small $r_0$ the final state consists mostly of light
quarks, whereas for larger systems the most abundant particle species
are pions. This can be explained with the shorter lifetime of the
smaller systems, which inhibits a complete conversion of all quarks
into mesons. As the system size becomes larger, the lifetime increases,
thus allowing more quarks to create hadrons \cite{bitch}.

It can also be seen in Tab.~\ref{multab1} that the final state contains
very few kaons and strange quarks; even in the largest system the ratio
of strange particles to all particles is about 1\%. The reason for this
is the large strange quark mass together with the $\Theta$-factor in
Eqs.~(\ref{plight}), (\ref{pstrange}). Due to the large strange quark
mass, a pair of a light quark and antiquark must have a large kinetic
energy in order to be able to create a strange-antistrange pair via
the process $u\bar u$, $d\bar d\to s\bar s$. Due to cutoff in the
initial conditions, on the other hand, such pairs do not occur very
frequently. The same mechanism inhibits the creation of kaons from light
quark pairs, since the kaon mass is of the same order of magnitude as
the strange quark mass. Furthermore, the cross section for the creation
of kaons from light quark pairs is small compared to the cross section for
their creation from strange quarks, so that kaons are most efficiently
produced from the latter \cite{su3hadron}. One observes, on the other
hand, a comparably large multiplicity of $\eta$ mesons, which on a first
glance might be surprising, since $\eta$ mesons have about the same mass
as kaons. A closer look at the production mechanisms shows however that
$\eta$ mesons are predominantly created by processes of the type $u\bar
u\to\pi^0\eta$ and variations hereof. This leads to a smaller mass
gap and thus to a larger $\eta$ than kaon production. We conclude
thus that it is easier to obtain a chemical equilibrium for $\eta$
mesons than for kaons, which might be a fact which persists to more
realistic models. Experimentally, this should be easy to verify once
the absolute multiplicities are known, since the $\eta$ has about the
same mass as the kaon, so that the multiplicity ratio in equilibrium
is given by the relative degrees of freedom: one has $N_\eta/N_K=1/4$
in chemical equilibrium, $N_\eta/N_K>1/4$ outside chemical equilibrium.

Since the multiplicity of strange particles cannot be reasonably
discussed using initial conditions containing a momentum cutoff,
we have also performed calculations with the $\Theta$-factors
in Eqs.~(\ref{plight}) and (\ref{pstrange}) dropped. While this
does not give a large effect on the particle masses, one has
more high momentum quark pairs at hand to create also heavier
particles. As already mentioned, this is paid with the loss of
exact energy conservation \cite{trap}, but we do not expect this
effect to have a large impact. The result for the multiplicities
is shown in Tab.~\ref{multab2}. As in Tab.~\ref{multab1}, it can
be seen that the ratio of mesons to quarks rises with $r_0$. The
ratio of strange particles to all particles stays approximately
constant at a level of ca.~10\%. The ratio $N_\eta/N_K$ is always
clearly above $1/4$, thus indicating that a complete chemical
equilibrium between these particles is not reached.

The evolution of the particle multiplicities with time for one of these
runs with $r_0=6$\,fm is shown in Fig.~\ref{multifig}.  It can be seen
here that the production of mesons starts already at the beginning of the
evolution. More details can be seen from Fig.~\ref{densev}, which shows
the angular averaged particle density as a function of $r$ for different
times. As one can see here, the meson density is high at those points,
where the quark density is high. Note however, that this is not true
in the very early stage of the expansion, where the temperature in the
centre is still high, as was discussed in Ref.~\cite{bitch}.

As a third variant of the initial conditions, we display in
Tab.~\ref{multab3} the particle multiplicities for the case that
one has strange quarks in the initial state, i.\,e. $f_s=1$ in
Eq.~(\ref{pstrange}). In this case, the ratio of strange particles to
the total number of particles lies around 20\%, which is consistent
with the values obtained in heavy ion experiments \cite{marek}. The
number of kaons is high compared to Tab.~\ref{multab1}, since now they
can be produced efficiently from the strange quarks which are present
in the initial state. In this case, the argument that an $\eta$ meson
can be produced more easily than a kaon remains no longer true, so that
the ratio $N_\eta/N_K$ is now lower than $1/4$. Thus the $N_\eta/N_K$
ratio may serve as a messager of the strange quark content at the
beginning of the expansion.

Comparing the multiplicities of Tabs.~\ref{multab1}--\ref{multab3},
we find as a consequence that we can only obtain a strangeness ratio
in the order of the experimentally observed one, if we assume that
strange quarks are already present in the initial state. In an
experimental situation, these might be created in collisions taking
place before the plasma is formed, which we do not treat here.

The time dependence of the multiplicities for the calculation for
$r_0=6$\,fm of Tab.~\ref{multab3} is displayed in Fig.~\ref{multif1}.
The kaon production according to this figure proceeds fast in the
beginning and terminates after 5\,fm$/c$. The pion production, on the
other hand, has two components: one fast component until $t=5$\,fm$/c$,
followed by a slow rise at later times, which finally saturates. This
time behaviour of the pion multiplicity is also found in calculations
which do not contain strange quarks initially, although this is not
shown explicitly here.  The second component of the pion production is,
however, not observed in Fig.~\ref{multifig}, where the quark momenta
have been extended up to high values.

In Fig.~\ref{thisto}, we show the creation probability of mesons as a
function of temperature. This figure was taken from a calculation with
an initial radius of 6\,fm and a momentum distribution without cutoff,
but the result does not strongly vary for different initial conditions,
except for statistical fluctuations. The solid curve, which denotes the
pion creation probability, shows the same behavior as for a two flavor
plasma \cite{bitch}. The pion production takes place at a temperature well
below the pion Mott temperature, within a temperature range of roughly
$150\,\mbox{MeV}<T<200\,\mbox{MeV}$. The same is true for kaons and $\eta$
mesons, for which the creation probability does not show a significant
deviation of the pion creation probability. The qualitative behaviour of
these curves can be understood by considering the mean hadronization time
of quarks, which for thermal equilibrium has been displayed in Fig.~23
of Ref.~\cite{su3hadron}. It has been shown there that this quantity
has a minimum for $150\,\mbox{MeV}<T<200\,\mbox{MeV}$, thus leading to
a minimum of the mean free path, so that one observes a maximum of
hadronization processes in this range. A qualitatively similar curve
as in Fig.~\ref{thisto} has been obtained in Ref.~\cite{bitch}, where,
however, only the pion case was considered.

The mean production temperatures according to Fig.~\ref{thisto} amount
to 168\,MeV for pions, 164\,MeV for kaons and 178\,MeV for $\eta$
mesons. These almost identical numbers have to be compared with the
Mott temperature, at which in an equilibrated, adiabatically expanding
system the hadronization takes place. Due to the finite mean free path,
the temperature in our calculations drops well below the Mott temperature
before the hadronization can take place.

\subsection{Expansion Dynamics}
In Fig.~\ref{rmsrad}, we show the root mean square distances of the
individual particle species from the centre for a system with initial
conditions as in Tab.~\ref{multab1} and initial radius 7\,fm. As can
be seen, the root mean square radius is largest for pions, followed
by light quarks, kaons, strange quarks and $\eta$ mesons. Thus the
mesons separate in coordinate space during their expansion, which
has consequences for the rescattering models on the hadronic level,
as employed by standard cascade calculations. The same picture can be
found in Tab.~\ref{wassolls}, where we give the mean energy, momentum
and velocity for each particle species in the final state. Again one
finds the largest velocity for pions, followed by light quarks, kaons,
$\eta$ mesons and strange quarks. This order is roughly an ordering by
the particle masses and can be explained partially by the production
mechanisms of the particles. If one considers e.\,g. pions, then one has
to take into account that pions are mainly produced via the annihilation
of light quark-antiquark pairs. For this process, the rest mass of the
final state is usually lower than the rest mass of the initial state,
so that the momenta of the pions have to be higher than those of the
quarks, which in turn leads to a higher velocity.  The same is true if
a kaon pair is generated from a strange-antistrange pair. During the
expansion, quarks are slowed down by the mean field, whereas mesons
become accelerated. This leads to a even higher momentum difference
between quarks and mesons. However, strange quarks are not so strongly
slowed down as light quarks, since their mass is already high in the
initial state. Although the above argument should be also true for
strange quarks, which are created from light quarks, we find thus a
slightly higher momentum for strange quarks as compared to light quarks.

Table~\ref{wassolls} shows that the complicated dynamics of the
expansion of the $SU(3)$ plasma creates an asymptotic momentum of the
kaons and pions which is for our initial condition close to experiment
at midrapidity \cite{na44}. We see as well that the mean momentum of
kaons is larger than the one of pions.  This effect, which has also
been observed by experiment, has been interpreted as a consequence of
rescattering of the mesons \cite{sorgen}. We see here that this increase
is not necessarily a hadronic effect, but may already be created during
the expansion of the plasma phase, if there is one. The meson momenta
are always larger than the momenta of the collision partners which have
produced them. This again reflects the fact that at the moment of their
creation the temperature is below the Mott temperature and therefore
the mass of the $q\bar q$ pair is considerably higher than that of the
two produced mesons.

\section{Summary and Conclusions} \label{sumsec}
In this paper we have studied the expansion and hadronization of
a quark-meson plasma using an effective chiral interaction. We have
concentrated on aspects of the evolution which are specific to the $SU(3)$
case. Other topics, which can be treated in the two flavor case as well as
in the three flavor case have been discussed in Ref.~\cite{bitch}.

We find that the strange quark mass is considerably larger than the
current strange quark mass and that it gets even enhanced by
initial conditions not containing strange quarks. This has consequences
for the production dynamics, since the threshold for the creation of a
strange quark pair rises. We find that it is difficult to produce the
experimentally observed ratio of strange particles to all particles
unless we assume the presence of strange quarks in the initial state.
In any case our findings question the assumption of thermal free gas
models, in which quarks are treated as particles with their bare mass,
or hydrodynamical calculations, which use an equation of state which is
based on this assumption.

Concerning the expansion dynamics, we find that the different
particle species become separated in coordinate as well as in momentum
space. Partially this can be understood by the production mechanisms,
as e.\,g. the relation between light quark and pion momenta, for
other species one has concurring effects, so that there is no simple
explanation. Using an initial density of 1--2\,GeV/fm$^3$, which has been
proposed for the phase prior to the expansion in a plasma created at SPS
energies, we find that the average momenta of pions and kaons observed
in our calculations are roughly consistent with experimental data. These
two effects cannot be modeled in hydrodynamical calculations as well.

The heavy ion experiments performed at CERN do not show a baryon free
($\mu = 0$) midrapidity region. To understand these reactions in detail,
the expansion of a plasma at finite baryochemical potential has to be
studied. This is presently under way.



\begin{table}
\begin{tabular}{c||c|c||c|c|c}
$r_0$ (fm) & $N_q$ & $N_s$ & $N_\pi$ & $N_K$ & $N_\eta$ \\
\hline
  3 &   144 &    0 &   81 &    2 &    1 \\
  4 &   303 &    1 &  215 &    1 &   10 \\
  5 &   541 &    5 &  477 &    5 &   20 \\
  6 &   881 &    5 &  879 &    7 &   40 \\
  7 &  1310 &   14 & 1497 &   18 &   57 \\
\end{tabular}
\caption[]{Particle Multiplicities in the final state. The numbers are
           summed over all internal degrees of freedom, such as flavor,
           particle/antiparticle and isospin. No strange particles are
           present in the initial state.}
\label{multab1}
\end{table}

\begin{table}
\begin{tabular}{c||c|c||c|c|c}
$r_0$ (fm) & $N_q$ & $N_s$ & $N_\pi$ & $N_K$ & $N_\eta$ \\
\hline
  3 &  399  &   21 &  224 &   37 &   27 \\
  4 &  897  &   61 &  560 &  117 &   41 \\
  5 & 1304  &  101 & 1185 &  171 &  118 \\
  6 & 2939  &  203 & 2186 &  291 &  154 \\
\end{tabular}
\caption[]{Particle Multiplicities in the final state for initial
           conditions without momentum cutoff. The numbers are
           summed over all internal degrees of freedom, such as flavor,
           particle/antiparticle and isospin. No strange particles are
           present in the initial state.}
\label{multab2}
\end{table}

\begin{table}
\begin{tabular}{c||c|c||c|c|c}
$r_0$ (fm) & $N_q$ & $N_s$ & $N_\pi$ & $N_K$ & $N_\eta$ \\
\hline
  3 &   118 &   36 &  117 &   30 &    3 \\
  4 &   303 &   87 &  244 &   71 &   11 \\
  5 &   557 &  147 &  539 &  143 &   16 \\
  6 &   862 &  234 & 1076 &  248 &   27 \\
\end{tabular}
\caption[]{Particle Multiplicities in the final state for initial
           conditions with strange quarks in the initial state. The
           momentum cutoff has been allied to the initial conditions. The
           numbers are summed over all internal degrees of freedom,
           such as flavor, particle/antiparticle and isospin.}
\label{multab3}
\end{table}

\begin{table}
\begin{tabular}{c||c|c|c|c}
       & $\left< E \right>$ (MeV) &  $m$ (MeV) & $p$ (MeV) & $v$ \\
\hline
   $q$ &        525               &       368  &     374   &   0.713 \\
   $s$ &        676               &       550  &     393   &   0.581 \\
 $\pi$ &        441               &       135  &     420   &   0.952 \\
   $K$ &        698               &       497  &     490   &   0.702 \\
$\eta$ &        674               &       515  &     435   &   0.645 \\
\end{tabular}
\caption[]{Mean particle energies, masses, momenta and velocities in
           the final state. The initial conditions are the same as in
           Table~\ref{multab1}, the initial radius is 7\,fm.}
\label{wassolls}
\end{table}


\begin{figure}
\caption[]{Feynman graph for the irreducible pseudoscalar polarization.
           Solid lines denote constituent quarks.}
\label{polarfig}
\end{figure}

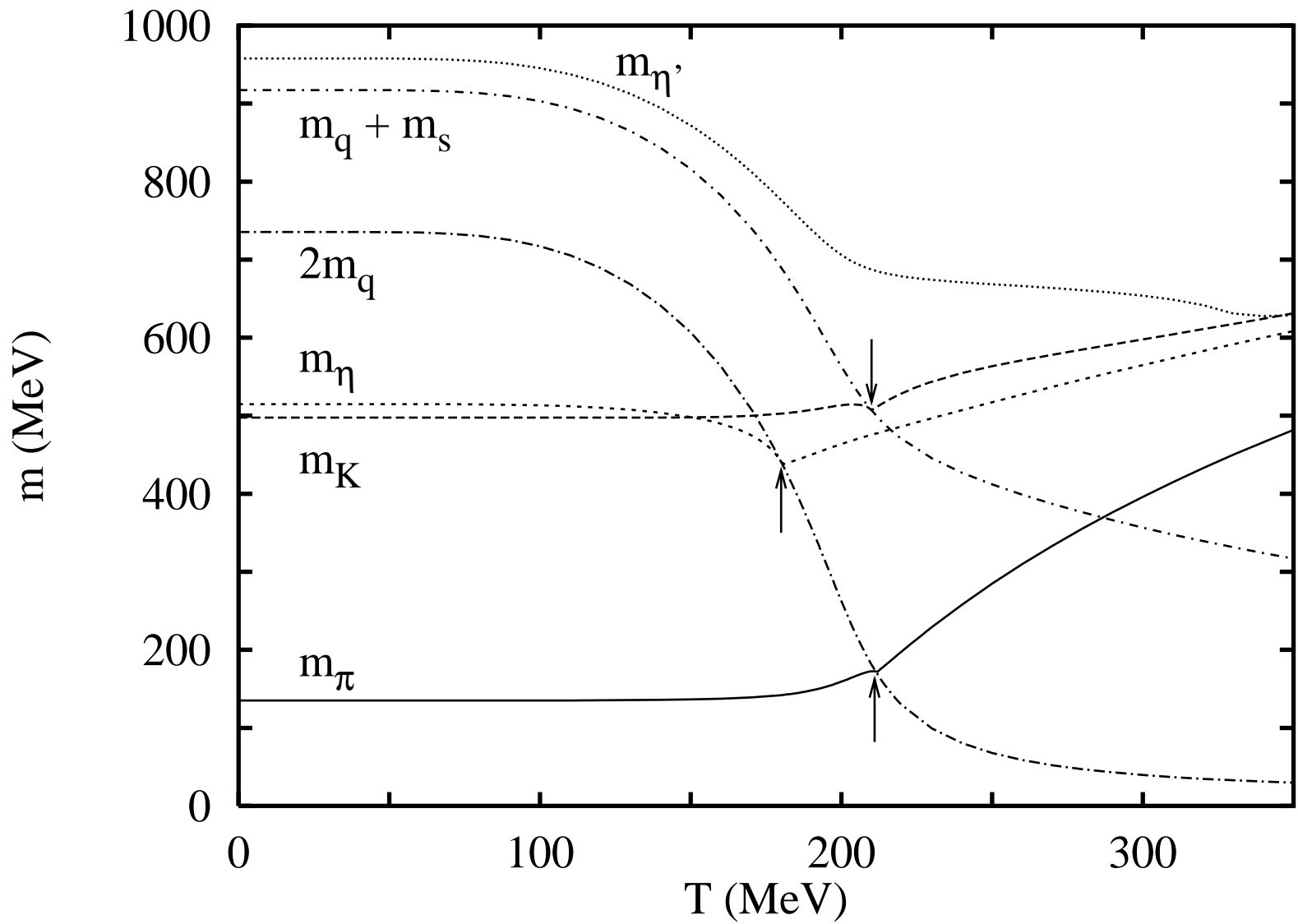
\begin{figure}
\caption[]{Meson masses as a function of temperature. Also shown are
           $2m_q$ and $m_q+m_s$. The Mott transitions of the pion, the
           kaon and the eta are marked by the arrows.}
\label{mmass}
\end{figure}

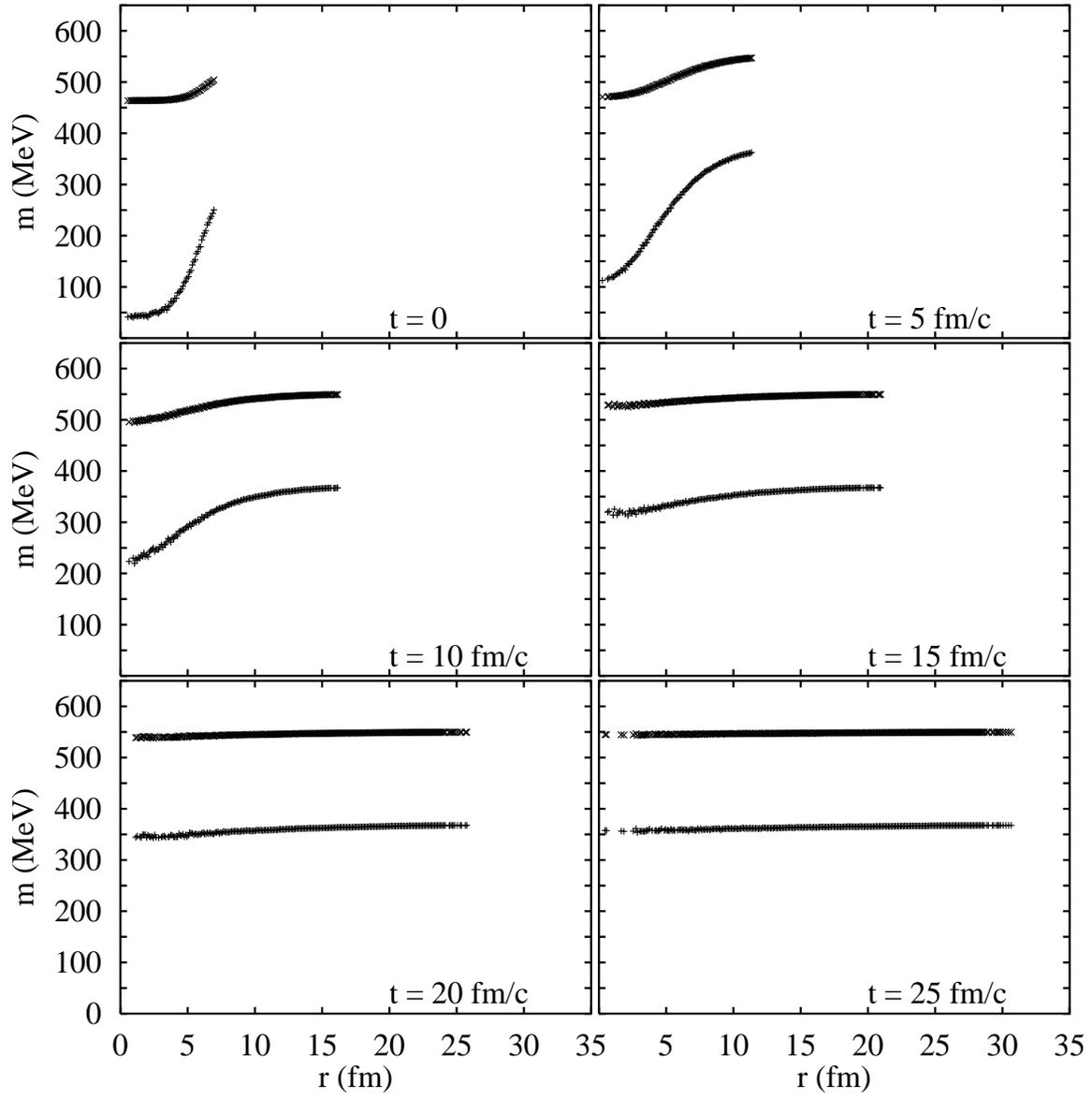
\begin{figure}
\caption[]{Constituent quark masses averaged over all solid angles as
           a function of $r$ for different times. The lower line is for
           light quarks, the upper line for strange quarks, respectively.}
\label{massev}
\end{figure}

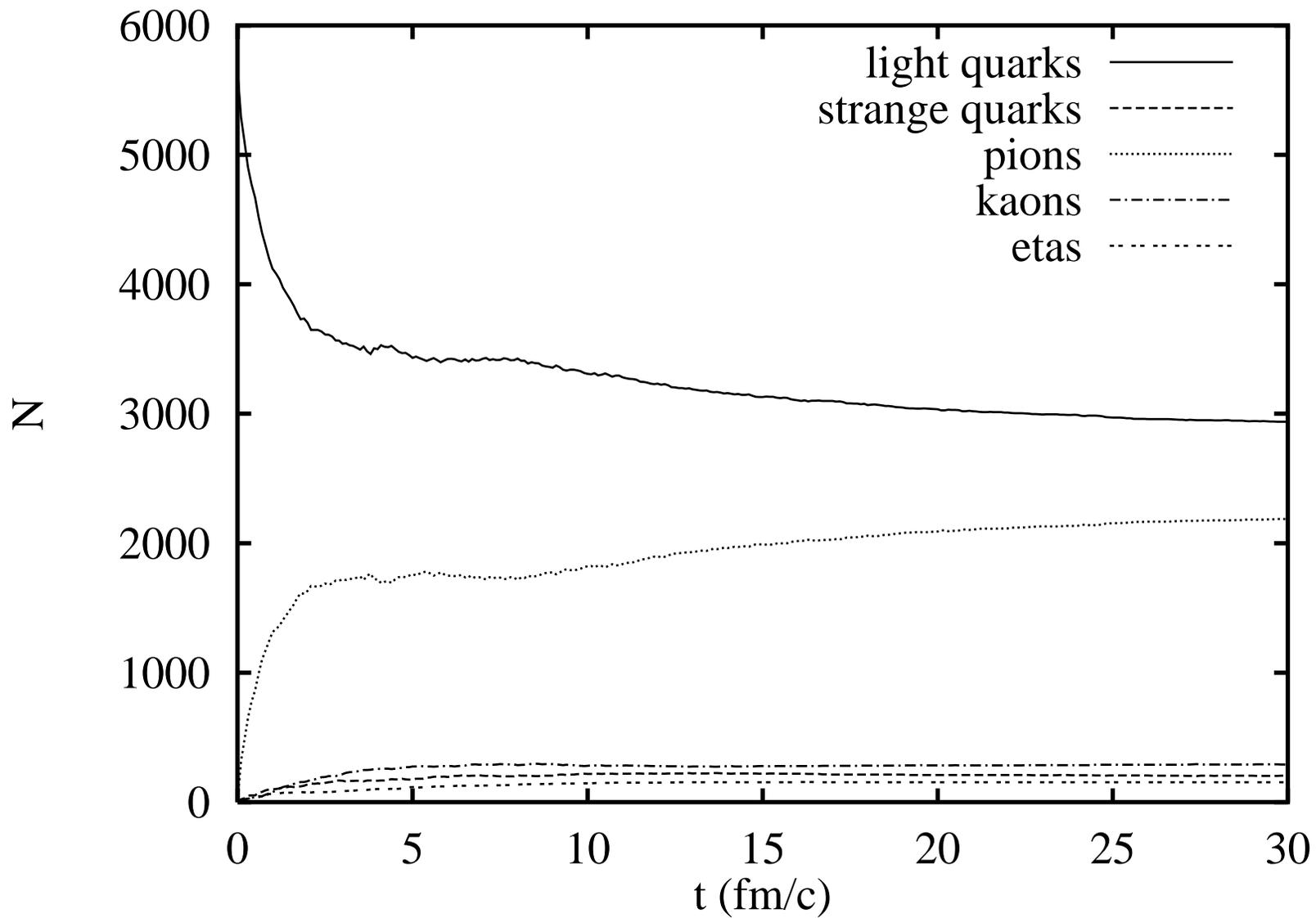
\begin{figure}
\caption[]{Time evolution of the particle multiplicities with initial
           condition corresponding to Table~\ref{multab2}. Solid line:
           light quarks, dashed line: strange quarks, dotted line: pions,
           dot-dashed line: kaons, double dashed line: $\eta$ mesons.}
\label{multifig}
\end{figure}

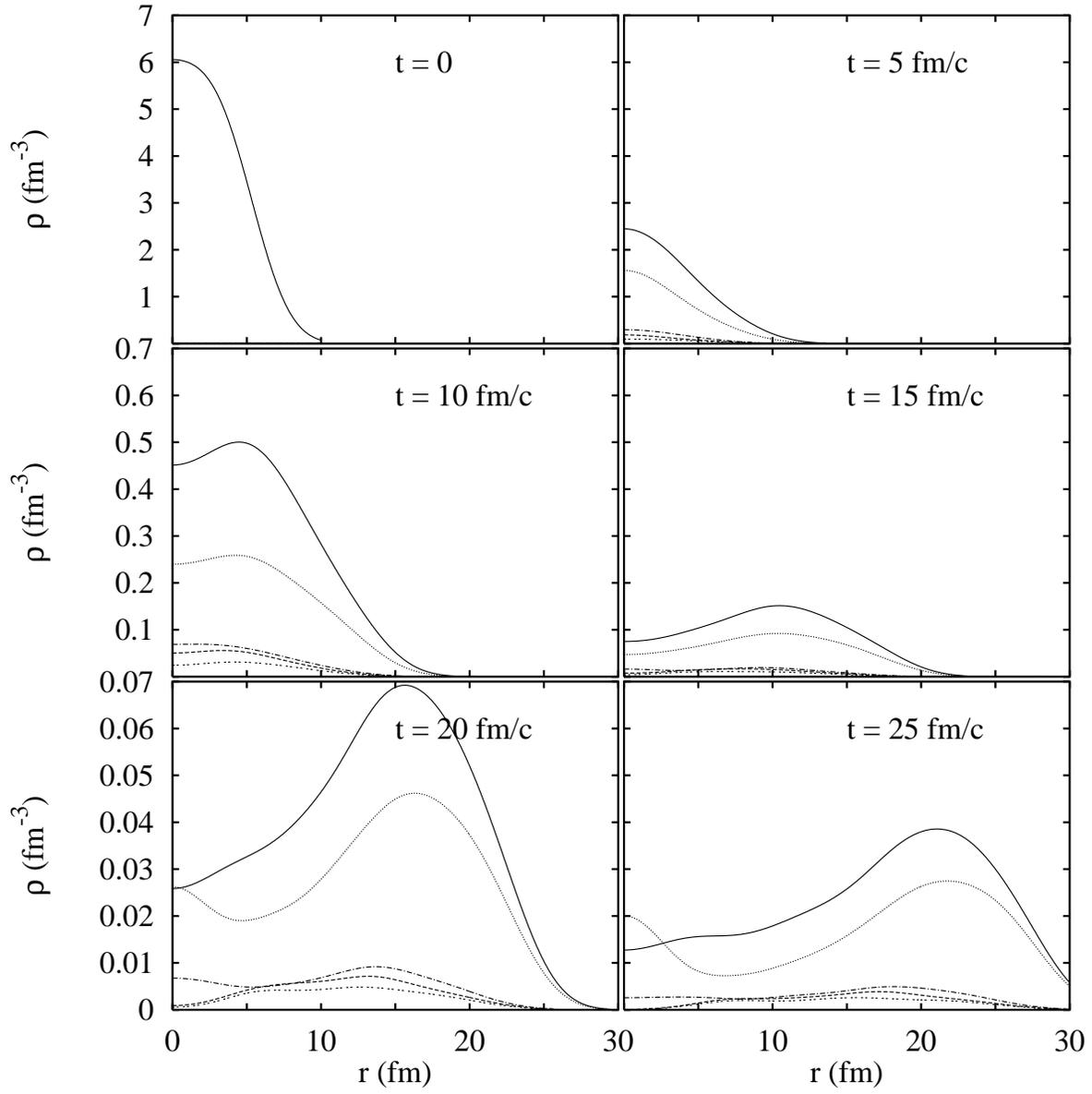
\begin{figure}
\caption[]{Time evolution of the angular averaged particle density
           with initial condition corresponding to Table~\ref{multab2}.
           Solid line: light quarks, dashed line: strange quarks, dotted
           line: pions, dot-dashed line: kaons, double dashed line: $\eta$
           mesons. Please note the different scales at the vertical axes.}
\label{densev}
\end{figure}

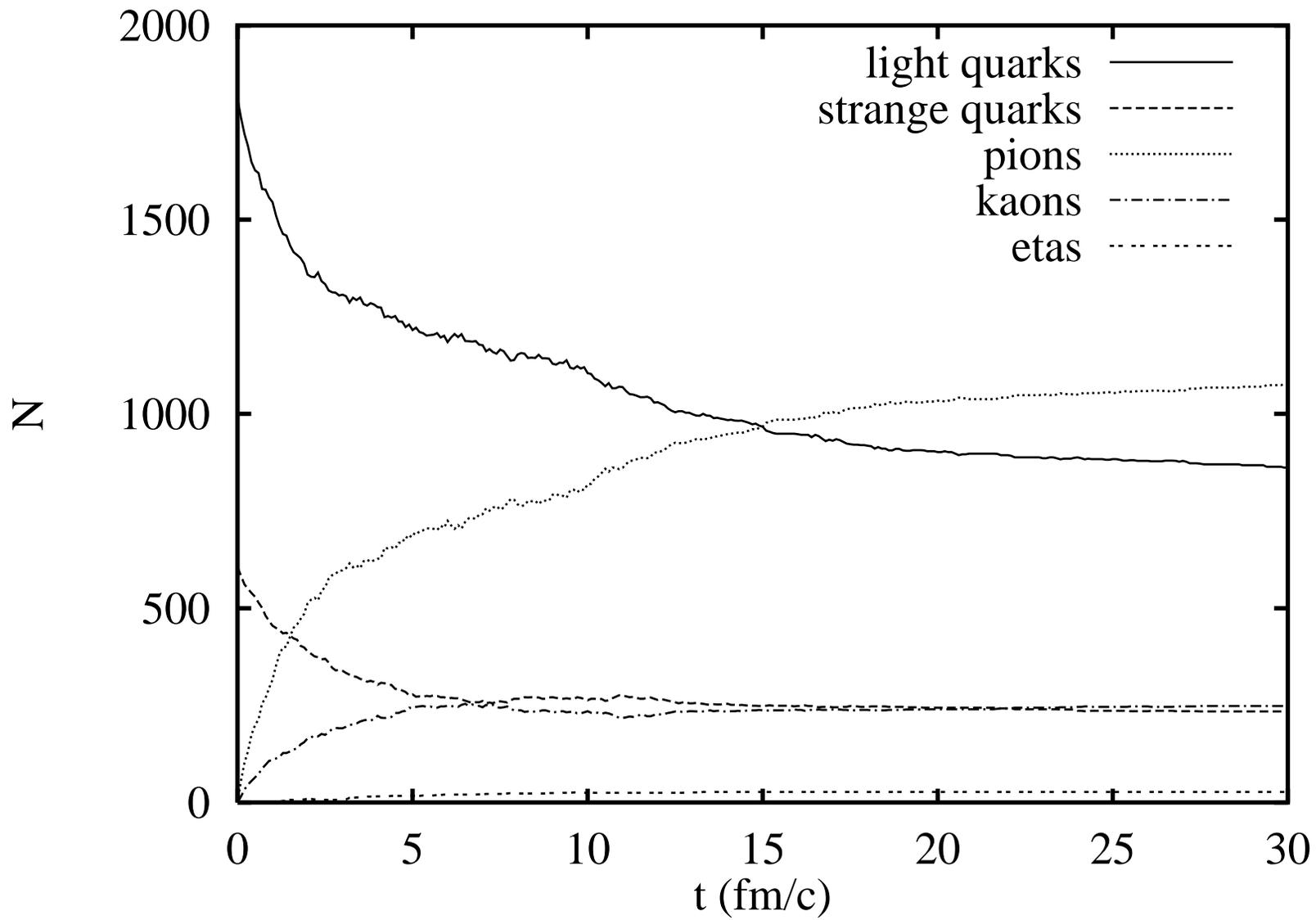
\begin{figure}
\caption[]{Time evolution of the particle multiplicities with initial
           condition corresponding to Table~\ref{multab3}. Solid line:
           light quarks, dashed line: strange quarks, dotted line: pions,
           dot-dashed line: kaons, double dashed line: $\eta$ mesons.}
\label{multif1}
\end{figure}

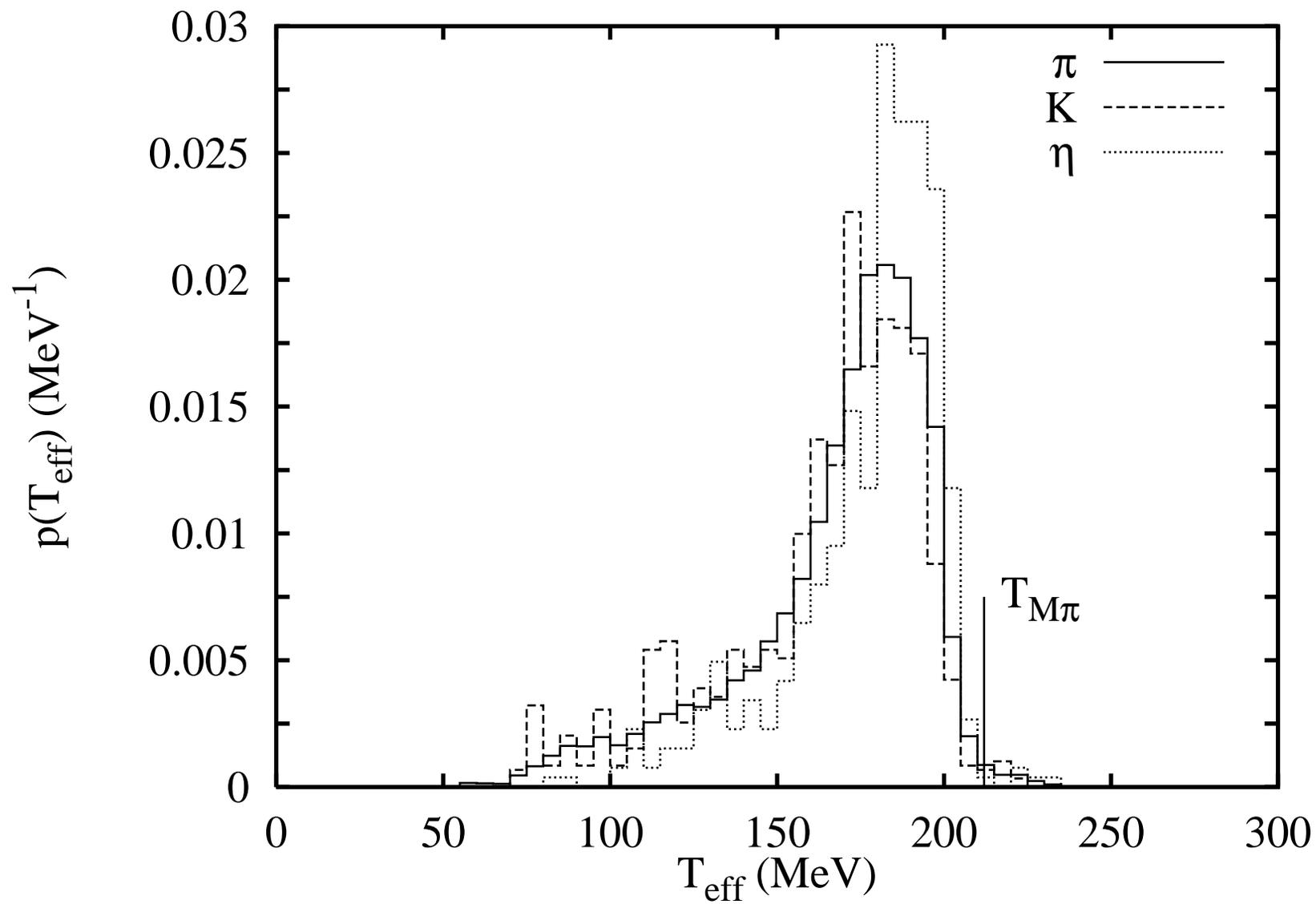
\begin{figure}
\caption[]{Creation probability of mesons as a function of temperature.
           Solid line: pions, dashed line: kaons, dotted line: $\eta$
           mesons.  The pion Mott temperature is indicated by the
           vertical line.}
\label{thisto}
\end{figure}

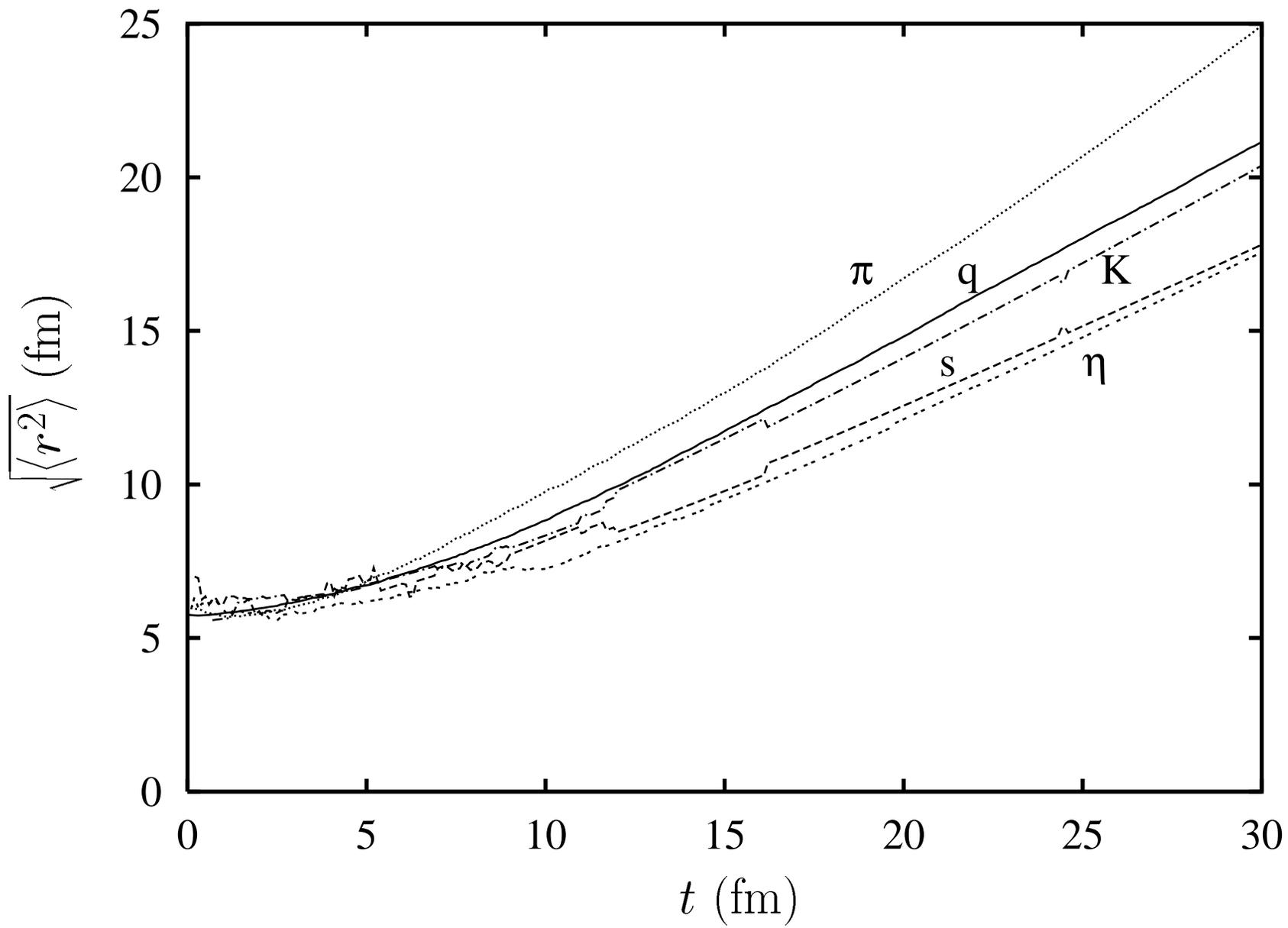
\begin{figure}
\caption[]{Root mean square radius for each particle species as a function
           of time. Solid line: light quarks, dashed line: strange
           quarks, dotted line: pions, dot-dashed line: kaons, double
           dashed line: $\eta$ mesons.}
\label{rmsrad}
\end{figure}

\end{document}